\documentclass[conference,a4paper]{IEEEtran}
\IEEEoverridecommandlockouts

\AddToHook{shipout/before}{%
  \ifnum\value{page}=3
    \setbox\ShipoutBox=\vbox{%
      \vskip0.024cm
      \box\ShipoutBox
    }%
  \fi
}
\usepackage{cite}
\usepackage{amsmath,amssymb,amsfonts}
\usepackage{graphicx}
\usepackage{textcomp}
\usepackage{xcolor}

\usepackage{amsmath, amssymb, amsthm,algorithm}
\usepackage[latin1]{inputenc}
\usepackage{xcolor}
\usepackage{latexsym}
\usepackage{graphics,epsfig}
\usepackage{amsfonts}
\usepackage{booktabs}
\usepackage{color}
\usepackage{multirow}
\usepackage[justification=centering]{caption}
\usepackage{graphicx}
\usepackage{adjustbox}
\usepackage{kantlipsum}
\usepackage[caption=false,font=footnotesize]{subfig}
\usepackage{cases}
\usepackage{url}
\usepackage{tabu}
\usepackage{makecell}
\usepackage{soul}
\usepackage{booktabs,tabularx}
\usepackage{array}
\usepackage{algpseudocode}
\usepackage[justification=centering]{caption}

\usepackage[caption=false,font=footnotesize]{subfig}
\usepackage[T1]{fontenc}
\usepackage{mwe}
\usepackage{subfig}
\usepackage{mathtools,lipsum,cuted}

\usepackage[english]{babel}
\usepackage{verbatim}

\usepackage[english]{babel}

\newtheorem{thm}{Theorem}

 \newtheorem{lemma}{Lemma}

\theoremstyle{remark}

\algrenewcommand{\algorithmiccomment}[1]{\hspace{-20px}$\rightarrow$ #1}

\def\BibTeX{{\rm B\kern-.05em{\sc i\kern-.025em b}\kern-.08em
    T\kern-.1667em\lower.7ex\hbox{E}\kern-.125emX}}
\begin{document}

\title{Throughput Analysis and On-Board Buffer Sizing for Hybrid RF and Optical LEO Satellites}
\author{\IEEEauthorblockN{Cao-Vien Phung, Thomas R\"othig, and Admela Jukan}
\IEEEauthorblockA{Institut f\"ur Datentechnik und Kommunikationsnetze, Technische Universit\"at Braunschweig, Germany\\
Email: \{c.phung, t.roethig, a.jukan\}@tu-bs.de}}
\maketitle

\begin{abstract}
Low Earth Orbit (LEO) satellite networks are increasingly adopting laser communications based on Free Space Optics (FSO). Although laser inter-satellite links offer high throughput and low latency, RF up- and downlinks remain necessary to maintain connectivity during optical outages caused by adverse atmospheric conditions. The hybrid RF/FSO up- and downlinks require an efficient transmission scheduling and buffer sizing in satellite network nodes, due to traffic asymmetry and variable capacity.  This paper analyzes throughput performance and buffer sizing in hybrid RF/laser satellite networks with finite buffer capacity, interference-aware scheduling, and weather-dependent laser link outage probabilities. Numerical results indicate that laser communications bring significant performance gains. Instead of increasing the transmission power of the satellite to maximize the throughput, we can select a suitable transmission scheduling priority to achieve a maximum throughput, while minimizing the buffer requirement, and lowering packet loss probability under varying weather conditions and constraints.
 \end{abstract}

\section{Introduction} \label{intro}
Modern satellite networks increasingly consider laser communication for optical up- and downlinks (LaserCom, Free Space Optics, FSO), as well as laser inter-satellite links (LISLs) ~\cite{Cardakli:26}. However, relying solely on laser links remains challenging: due to weather effects, including clouds, fog, and turbulence, optical up- and downlink performance can deteriorate. In addition, optical wavelength reuse causes co-channel interference~\cite{9203155}. Therefore, hybrid FSO/RF links still remain necessary ~\cite{9655260}. Although  buffering can reduce packet loss during link outages and contention~\cite{9961134}, onboard processing and memory resources remain limited by satellite SWaP (size, weight, power) budgets, making arbitrarily large buffers impractical at Gbps transmission~\cite{9961134}.

We envision that future all-optical satellite networks can incorporate transparent all-optical forwarding lowering latency, and improving scalability. Ideally, such networks would operate with small or no buffering. For now, fully bufferless operation is however not feasible in practice, most notably because optical feeder links may be intermittently unavailable due to weather effects.  Higher transmit power can reduce the optical link outage probability~\cite{9655260}, thereby lowering required onboard buffering. This maybe practical at ground stations, but it can still be impractical on satellites due to SWaP constraints ~\cite{7124715}.  Before a fully bufferless all-optical satellite networking becomes feasible, a practical intermediate step is optimize onboard buffering and throughput performance.

This paper investigates the minimum onboard buffer capacity required to maintain a certain level of throughput performance under various weather conditions affecting optical satellite up- and downlinks. We propose a Markov chain model for the throughput analysis and onboard buffer sizing. The model characterizes system performance as a function of finite satellite buffer capacity, transmission scheduling parameter, and link outage probabilities under varying weather conditions, and it optimizes transmission scheduling. Numerical results show that instead of increasing the transmission power of the satellite to maximize the performance, as it is commonly the case, we can select a suitable optical transmission scheduling priority to achieve maximum throughput with minimum buffer size, while lowering the packet dropping probability.


The rest of this paper is organized as follows. Section \ref{rewkfia} presents the related work. Section \ref{buffanablockprob} provides the buffer and throughput analysis.  Section \ref{perforevablockedfnm} presents performance evaluation. We conclude the paper in Section \ref{conclpaper}. 

\section{Related work} \label{rewkfia}
Paper \cite{9655260} considers satellite communications under different weather conditions and it analyzes the link outage probability using a weather-dependent hybrid RF/laser transmission. We use the same weather scenarios and the corresponding outage probabilities, while extending it with interference-avoiding transmission scheduling. In terms of interference avoidance, paper \cite{CHI2014101} analyzes the interference-avoiding transmission scheduling in a two-hop indoor wireless network using a Markov chain model. While also using a Markov chain model, we extend this work, however considering outdoor satellite links and their outage. Paper \cite{11124522} also analyzes the performance of interference-avoiding transmission scheduling in satellite-aerial-ground integrated networks (SAGINs) with a finite relay buffer using a multidimensional Markov chain model.  We extend this approach by studying link outage caused by the varying weather conditions. Paper \cite{9217992}  develops an analytical model for mixed laser/RF UAV-aided mobile relaying with a finite buffer under foggy weather conditions. We extend this work by incorporating outage links for both inter-satellite and satellite to ground station links. In terms of the relationship between traffic and buffer size, papers  \cite{8717569,9961134} study satellite networks with limited buffer/cache, as they are used to re-transmit error/lost bursts, also called traffic chunks. These studies focus on selecting a data burst, or chunk size, transferred during its burst/chunk time, in relationship to the buffer size, which is different from our study. It should be noted that we mention the above related later in the analysis every time the specific aspects of the models are re-used. 

The main novelty of our work is the single--model analytical treatment of weather--dependent link availability, interference--aware transmission scheduling, and finite on-board buffering in a hybrid RF/FSO LEO satellite network, which other papers address in isolation. We jointly characterize the  throughput, buffer size, and packet dropping probability as functions of the transmission scheduling priority, buffer capacity, link outage probabilities, and weather conditions. Finally, we not only analytically derive the performance under individual weather conditions, but also in case of a likelihood of a certain weather probability over a period of time, thereby providing a multi-weather framework, which is novel. 
links. 

\begin{figure}[!t]
    \centering
    \includegraphics[width=0.94\linewidth]{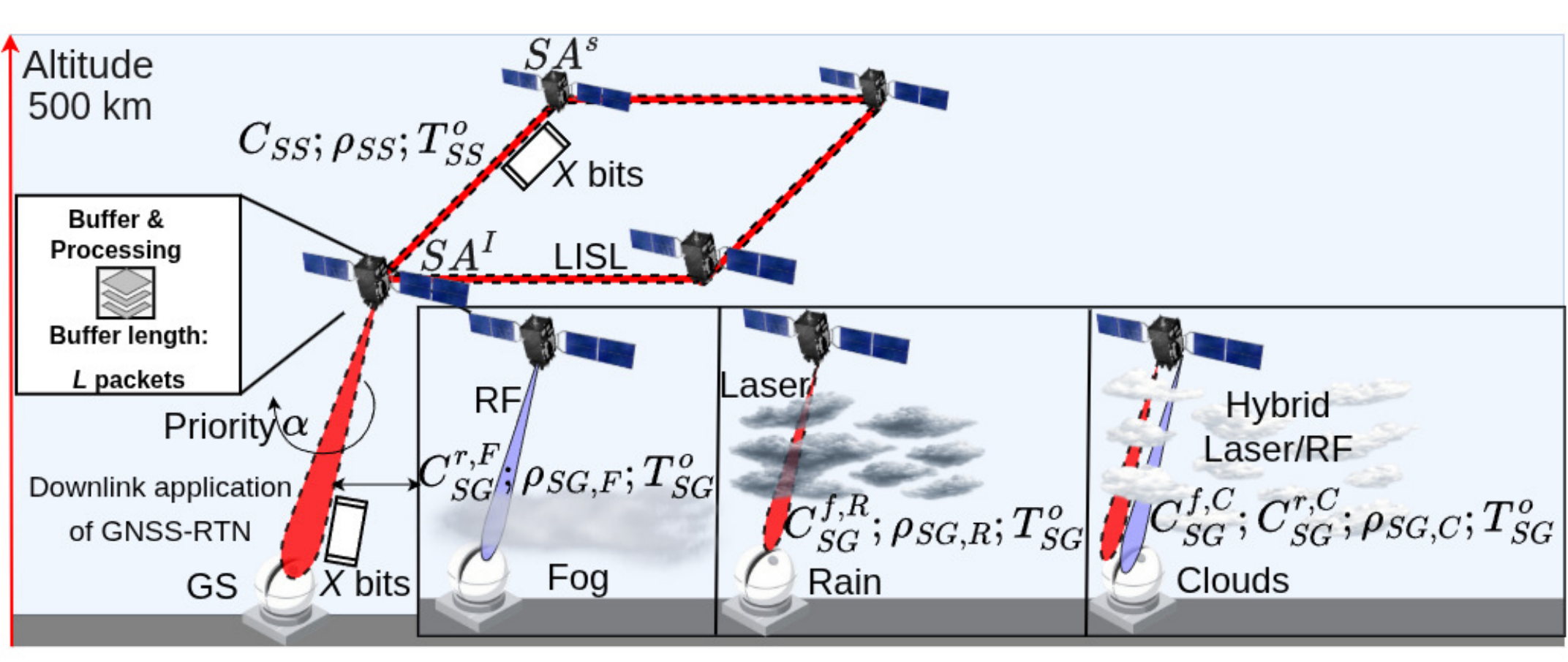}
    \caption{Reference scenario and different weather effects.}
    \label{scenario_blocked}
\end{figure}

\begin{table}[ht]
  \centering
  \caption{List of main parameters.}
  \label{listnotationblocked}
  \begin{tabular}{ll}
    \toprule
    \textbf{Notation} & \textbf{Meaning}\\
    \midrule
    $A_{\gamma}$ & Average queue length [bits],\\
                 & $A_{\gamma}=\{A_C: cloud, A_F: fog, A_R: rain \}$.\\
    $C_{\gamma}$ & Capacity [bps], $C_{\gamma} = \{C_{SG}^{f,C}: cloud \; with \; laser, C_{SG}^{f,R}:$ \\ 
    & $rain, C_{SG}^{r,C}: cloud \; with \; RF,  C_{SG}^{r,F}: fog, C_{SS}: LISL\}$.\\
    $L$, $L_b$ & Buffer length limit of satellite SA$^I$, $L$ [packets], $L_b$ [bits]. \\
   $O_t$ & Total time observed in the weather using RF link [seconds].\\
   $P_{\gamma}$ & Average packet dropping probability,\\
    & $P_{\gamma} = \{ P_C: cloud, P_F: fog, P_R: rain\}$.\\
    $P(i)$ & State probability of $i$ packets in the buffer.\\
    $P_l$ & Transition probability from state $i \in (0,L)$ to itself.\\
    $P_l^{'}$ & Transition probability from state $i=L$ to itself.\\
    $P_{SA^I}$ & Transition probability from state $i \in (0,L]$ to state $i-1$.\\
    $P_{SA^I}^{'}$ & Probability that SA$^I$ sends unsuccessfully data to GS.\\
    $P_{SA^s}$ & Transition probability from state $i \in (0,L)$ to state $i+1$.\\
    $P_{SA^s}^{'}$ & Probability that SA$^s$ sends unsuccessfully data to SA$^I$.\\ 
    $T_{SG}^o$ & Timeout for transmission via SA$^I$-GS link [seconds].\\
    $T_{SS}^o$ & Timeout for transmission via SA$^s$-SA$^I$ link [seconds].\\
    $X$ & Average size of a packet [bits]. \\
    $W_{\gamma}$ & Probability $W_{\gamma}=\{W_C: cloud, W_R: rain, W_F:fog \}$.\\
    $\alpha$ & SA$^I$ priority of accessing medium (real value).\\
    $\rho_{\gamma}$ & Outage probability, $\rho_{\gamma}=\{\rho_{SG,C}: cloud,$\\
    &  $\rho_{SG,F}: fog, \rho_{SG,R}: rain, \rho_{SS}: LISL\}$.\\
    $\tau_{\gamma}$ & Average throughput [bps],\\
    & $\tau_{\gamma}=\{\tau_C: cloud,  \tau_F: fog, \tau_R: rain\}$.\\
    \bottomrule
  \end{tabular}
\end{table}

\section{Buffer and throughput analysis} \label{buffanablockprob}
\subsection{Reference scenario} \label{assumptionsblocked}

  Our reference scenario in Fig. \ref{scenario_blocked} includes one LEO Satellite source (SA$^s$), one intermediate LEO Satellite (SA$^I$), and one GS (destination node), running for instance a Global Navigation Satellite System Real-Time Network (GNSS-RTN) for autonomous vehicle control with geospatial data retrieved in real time \cite{10663454}. SA$^I$ always connects with SA$^s$, either directly or by handover, under the assumption of a satellite grid. SA$^I$ can serve one or more GSs during its movements. For simplicity and without loss of generality, assume that SA$^I$ can connect with one SA$^s$ and one GS only.  Here, SA$^s$ is assumed to always have packets available for transmission, whereas SA$^I$ does not generate packets itself, as it only relays packets received from SA$^s$.  Table \ref{listnotationblocked} summarizes the notations.

\begin{table}[ht]%
\caption{Link selections between SA$^I$ and GS \cite{9655260}.}
\label{seleclinksatcom}
\begin{center}
\setlength\tabcolsep{3pt}
\begin{tabular}{|c|c|}
  \hline
\textbf{Weather condition} & \textbf{SA$^I$ to GS link} \\
\hline
Thin cloud & Both laser and RF links (dual channels) \\
\hline
Rain &  laser link \\
\hline
Fog & RF link \\
\hline
\end{tabular}
\end{center}
\end{table}


With regard to LISLs, as the SA$^s$-SA$^I$ laser link may experience unstable optical ISLs and outage events due to pointing errors, rapidly changing network topologies, intermittent connections, and other factors, packets transmitted from SA$^s$ may fail to reach SA$^I$ within the defined timeout interval $T^o_{SS}$. Let $\rho_{SS}$ denote the outage probability of the considered SA$^s$-SA$^I$ laser link. After the timeout interval, an unsuccessful packet due to an outage of the SA$^s$-SA$^I$ laser link can be retransmitted at the data link layer, as conceptualized in \cite{9961134}. The capacity of the SA$^s$-SA$^I$ LISL is given by \cite{10673969}:
\begin{equation} \label{csskia}
C_{SS} = W_f \cdot log_2(1+SNR_{SS}),
\end{equation}
where $W_f$ is bandwidth, and $SNR_{SS}$ is signal-to-noise ratio. 


The weather effects are summarized in Table \ref{seleclinksatcom}. For thin cloud, the identical rate information is transmitted by  both RF and laser channels in parallel\cite{9655260}, where the selection technique is to combine received signals at GS to maximize performance. For simplicity and without loss of generality, we assume that there exists only one link (either laser or RF) whose performance is calculated on average between laser and RF. Note that due to optical interference in the same frequency band \cite{9203155}, transmitting data to GS and receiving data from SA$^s$ at SA$^I$ for  weather conditions using SA$^I$-GS laser links cannot be simultaneously performed, whereas SA$^I$ can simultaneously transmit data to GS for weather conditions using RF and receive data from SA$^s$ using laser. 

Regarding the SA$^I$-GS laser/RF links, they may also experience unstable link conditions and outage events due to weather effects, such as turbulence, clouds, rain, and fog. Consequently, packets transmitted from SA$^I$ may fail to reach the GS within the defined timeout interval $T^o_{SG}$. Let $\rho_{SG,C}$, $\rho_{SG,R}$, and $\rho_{SG,F}$ denote the outage probabilities of the SA$^I$-GS links under thin cloud-, rain-, and fog conditions, respectively, as in \cite{9655260}. After the timeout interval, a packet that fails to reach the GS due to an outage of the SA$^I$-GS laser/RF link can be retransmitted at the data-link layer, as conceptualized in \cite{9961134}.  SA$^I$-GS link capacities $C_{SG}^{f,C}$, $C_{SG}^{f,R}$, $C_{SG}^{r,C}$, and $C_{SG}^{r,F}$ correspond to thin cloud conditions with the laser link, rain,  thin cloud with the RF link, and fog conditions, respectively. These capacities are calculated similarly to Eq. \eqref{csskia}, with the corresponding SNRs obtained from \cite{9655260}. We assume that the switching time between the laser and RF links, including the signal-processing time at SA$^I$, is negligible.

With regard to the SA$^I$ buffer, it has a capacity of $L$ packets ($L_b$ bits). First, regarding transmission scheduling, when  SA$^I$-GS laser link is used, SA$^s$ is scheduled using Time-Division Multiplexing (TDM) to transmit one packet to SA$^I$ in each transmission opportunity, where all exchanged packets have the same size of $X$ bits. Each packet is stored in the buffer and occupies one buffer unit until its scheduled transmission opportunity from SA$^I$ to GS. With the TDM method, we define $\alpha$ "priority" (real value) for accessing the transmission medium from SA$^I$ to GS, i.e., $\alpha$ times of transmission opportunities from SA$^I$ to GS, whereas SA$^s$ competes to get one time for accessing the transmission medium from SA$^s$ to SA$^I$. SA$^I$ and SA$^s$ do not compete for access to the transmission medium when SA$^I$ buffer is empty. If the number of packets stored in the buffer reaches $L$, SA$^I$ drops any newly arriving packet. When SA$^I$ is granted a transmission slot, it transmits one packet to GS and removes that packet from its buffer. Second, regarding SA$^I$-GS link outage events, the SA$^I$ buffer temporarily stores packets that are unsuccessfully transmitted to GS. These packets are retained in the buffer and are released only when they are successfully received by GS. Finally, regarding the imbalance between incoming and outgoing transmission rates, which occurs under weather conditions in which SA$^I$-GS RF link is used, SA$^I$ does not apply the priority parameter $\alpha$ because the SA$^s$-SA$^I$ links are optical and therefore do not interfere with the SA$^I$-GS RF link. However, SA$^I$ still requires a buffer to temporarily store packets when there is a large imbalance between the high-rate incoming laser link and the low-rate outgoing RF link.



\subsection{Thin cloud weather model}

\subsubsection{Queue length analysis (thin cloud)}
\begin{lemma} \label{Actotalmorning}
The average queue length of the SA$^I$ buffer in the thin cloud using SA$^I$-GS laser/RF link can be given as follows:
\begin{equation}
A_C = \frac{1}{2}(A_{C,f}+A_{C,r}),
\end{equation}
where $A_{C,f}$ and $A_{C,r}$ are the average queue length of SA$^I$ if using SA$^I$-GS laser and SA$^I$-GS RF, respectively. $A_{C,f}$ and $A_{C,r}$ are calculated by using Lemmas \ref{theoremaveragequeuehaps} and \ref{Acrnamli}, respectively.
\end{lemma}

\begin{figure}[h]
    \centering
    \includegraphics[width=0.5\textwidth]{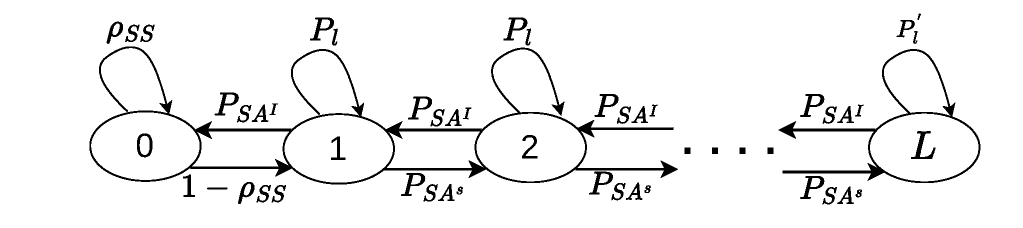}
    \caption{Embedded Markov chain model applied.}
    \label{Markov_block_prob}
\end{figure}

With cloud (SA$^I$-GS laser), we design a Markov chain (Fig. \ref{Markov_block_prob})  to obtain the throughput, queue length of buffered packets, and packet dropping probability at SA$^I$, where each state $i \in [0,L]$ is the number of buffered packets in the SA$^I$ queue (queue state). Observations in the Markov chain occur before each packet transmission from SA$^s$ or SA$^I$.

Let $P_{0 \rightarrow 1}$ and $P_{0 \rightarrow 0}$ be the transition probability from state $i=0$ (empty buffer) to state $i=1$ and to itself, respectively. Since the SA$^I$ buffer is empty, the transmission is possible only from SA$^s$, which increases by one unit in the queue. We get $P_{0 \rightarrow 1}=1-\rho_{SS}$ in case SA$^s$ sends successfully one packet to SA$^I$, whereas we get $P_{0 \rightarrow 0}=\rho_{SS}$ in case SA$^s$ sends unsuccessfully one due to SA$^s$-SA$^I$ link outage.

 Let $P_{SA^s}$ denote the transition probability that a nonempty queue in state $i$ $(0 < i < L)$ transits to state $i+1$, i.e., the queue length increases by one, when SA$^s$ is occupied and transmission from SA$^s$ is possible. As  SA$^I$ buffer is nonempty, SA$^I$ has transmission scheduling priority of $\alpha$ times over SA$^s$ when competing for transmission opportunities. Hence, given that SA$^s$ is occupied, the transition probability $P_{SA^s}$ that a nonempty SA$^I$ queue increases by one packet, i.e., SA$^s$ successfully transmits one packet to SA$^I$, is determined by SA$^s$-SA$^I$ link outage probability $\rho_{SS}$ and is given as follows:
\begin{equation} \label{eqsavienminhne}
P_{SA^s} = \frac{1-\rho_{SS}}{1+\alpha},
\end{equation}
whereas the probability $P_{SA^s}^{'}$ that the nonempty SA$^I$ queue stays unchanged (SA$^s$ sends unsuccessfully one to SA$^I$) is:
\begin{equation} \label{psacommablocked}
P_{SA^s}^{'} = \frac{\rho_{SS}}{1+\alpha}.
\end{equation}

Let $P_{SA^I}$ be the transition probability that nonempty queue state $i \in (0,L]$ to $i-1$, i.e., queue decreases one packet, in case of transmission chance from SA$^I$ occupied with the "$\alpha$ priority". With $\rho_{SG,C}$ of SA$^I$-GS laser link outage probability, the probability $P_{SA^I}$ that the nonempty SA$^I$ queue decreases one packet (SA$^I$ sends successfully one to GS) is given by:
\begin{equation} \label{hapseqhangphamne}
P_{SA^I} = \frac{\alpha(1-\rho_{SG,C})}{1+\alpha},
\end{equation}
whereas the probability $P_{SA^I}^{'}$ that the nonempty SA$^I$ queue keeps unchanged (SA$^I$ sends unsuccessfully one to GS) is:
\begin{equation} \label{phapscommablocked}
\small
P_{SA^I}^{'} = \frac{\alpha \cdot \rho_{SG,C}}{1+\alpha}.
\end{equation}

With Eqs. \eqref{psacommablocked}, \eqref{phapscommablocked}, the transition probability $P_l$ from state $i \in (0,L)$ to itself, i.e., the probability that the nonempty SA$^I$ queue stays unchanged (SA$^s$ or SA$^I$ sends unsuccessfully one packet to SA$^I$ or GS, respectively) is given as follows:
\begin{equation}  \label{plblockedsumsum}
P_l = P_{SA^s}^{'} + P_{SA^I}^{'} = \frac{\rho_{SS} + \alpha \cdot \rho_{SG,C} }{1+\alpha}.
\end{equation}

At $i=L$, any new packet that arrives at SA$^I$ when SA$^s$ sends data with the probability $P_{SA^s}$ will be dropped. With Eqs. \eqref{psacommablocked}, \eqref{phapscommablocked}, \eqref{plblockedsumsum}, the transition probability from $i=L$ to itself is given as follows:
\begin{equation}
P_l^{'} = P_{SA^s} + P_l= \frac{1+\alpha \cdot \rho_{SG,C}}{1+\alpha}.
\end{equation}

\begin{lemma} \label{fiablockednew}
The Markov chain is finite, irreducible, aperiodic.
\end{lemma}
The proof is provided in Appendix \ref{prooffiablockednew}. 

The Markov chain satisfies Lemma \ref{fiablockednew}, which has a steady-state distribution to which the distribution converges.
\begin{lemma} \label{stateprobblockednewvi}
Let $P(i)$ be the state probability that the SA$^I$ buffer has $i$ packets. $P(i)$ in thin cloud (SA$^I$-GS laser link) is:
\begin{equation}
P(0) = \frac{1}{ 1+(1-\rho_{SS})\sum_{i=1}^L \frac{(P_{SA^s})^{i-1}}{(P_{SA^I})^i} },
\end{equation}
\begin{equation} \label{varianteqblocked}
P(i) = \frac{ (P_{SA^s})^{i-1} }{ (P_{SA^I})^{i} } (1-\rho_{SS}) P(0), \forall i \in [1,L].
\end{equation}
\end{lemma}
The proof is provided in Appendix \ref{proofstateprobblockednewvi}.

\begin{lemma} \label{theoremaveragequeuehaps}
The average queue length of SA$^I$ buffer in the thin cloud using SA$^I$-GS laser link can be given as follows:
\begin{equation}
A_{C,f} = X(1-\rho_{SS}) P(0)  \sum_{i=1}^L i \frac{ (P_{SA^s})^{i-1} }{ (P_{SA^I})^{i} },
\end{equation}
where $P(0)$ is referred to Lemma \ref{stateprobblockednewvi}.
\end{lemma}
Lemma \ref{theoremaveragequeuehaps} proves by using $X \sum_{i=1}^L i \cdot P(i)$ and Lemma \ref{stateprobblockednewvi}.

\begin{lemma} \label{Acrnamli}
The queue length of SA$^I$ buffer (i.e., using SA$^I$-GS RF) is the difference between the capacities of SA$^s$-SA$^I$ laser and RF links with observed time $O_t$ s, is given by, i.e., 
\begin{equation}
 \begin{cases}
 \begin{split}
 & A_{C,r}  = \frac{1}{O_t+1} \sum_{i=0}^{O_t} min \left(L_b,i \cdot [\beta-\phi] \right),\text{if } \beta > \phi,\\
  & A_{C,r} = 0, \; \text{if } \beta \leq \phi, 
  \end{split}
  \end{cases} 
\end{equation}
where  $\beta=(1-\rho_{SS})\cdot C_{SS}$ represents the amount of data arriving at SA$^I$ from SA$^s$ per second and $\phi=(1-\rho_{SG,C})\cdot C^{r,C}_{SG}$ represents the amount of data released from SA$^I$ to GS per second.  Since the queue length of SA$^I$ buffer is determined only by the difference between the capacities of SA$^s$-SA$^I$ laser and RF links over the observation time $O_t$ s, specifying a packet size is unnecessary. 
\end{lemma}

\subsubsection{Packet dropping probability analysis (thin cloud)}
We define the packet dropping probability to be packets that arrives at SA$^I$ where refuses buffering them due to full buffer.
\begin{lemma} \label{pcquametroi}
The packet dropping probability at the SA$^I$ buffer in the thin cloud using SA$^I$-GS laser/RF link is given by:
\begin{equation}
P_C = \frac{1}{2}(P_{C,f} + P_{C,r}),
\end{equation}
where $P_{C,f}$ and $P_{C,r}$ are the packet dropping probability at SA$^I$ buffer in the thin cloud if using SA$^I$-GS laser and RF links, respectively. $P_{C,f}$ and $P_{C,r}$ are from Lemmas \ref{pdpblockedvido}, \ref{PCrmorning}.
\end{lemma}

\begin{lemma} \label{pdpblockedvido}
The packet dropping probability at SA$^I$ in the thin cloud using SA$^I$-GS laser link is given as follows:
\begin{equation}
P_{C,f} = \left( \frac{ P_{SA^s} }{ P_{SA^I} } \right)^{L} (1-\rho_{SS}) P(0),
\end{equation}
where $P(0)$ is referred to Lemma \ref{stateprobblockednewvi}.
\end{lemma}
Lemma \ref{pdpblockedvido} is proved by using $P_{SA} \cdot P(L)$ and Lemma \ref{stateprobblockednewvi}.

\begin{lemma} \label{PCrmorning}
The packet dropping probability at SA$^I$ in the thin cloud using SA$^I$-GS RF is given as follows:
\begin{equation}\label{xifogrd}
 \begin{split}
 P_{C,r} = 1-\left[ \frac{1}{O_t}  \sum_{i=1}^{O_t} min \left(1, \frac{S^{z}_e + \phi }{ \beta} \right) \right],
 \end{split}
\end{equation}
where the remaining SA$^I$ buffer space in bits at the observed time $z=i-1$ (previously observed time $i$) is given as follows:
\begin{equation} \label{remain_space}
 \begin{cases}
 \begin{split}
 & S_{e}^{z}  = L_b - min \left(L_b,z \cdot [\beta - \phi] \right), \text{if } \beta >  \phi, \\
  & S_{e}^{z} = L_b, \; \text{if } \beta \leq \phi,
  \end{split}
  \end{cases} 
\end{equation}
\end{lemma}
All parameters in Lemma \ref{PCrmorning} can be referred to Lemma \ref{Acrnamli}.

\subsubsection{Throughput analysis}
We define the average throughput to be the ratio of the average amount of data per transmission slot that SA$^I$ forwards to the average transmission slot length.
\begin{lemma} \label{taucafternoon}
The average throughput in the thin cloud using SA$^I$-GS laser/RF link can be given as follows:
\begin{equation}
\tau_{C} = \frac{1}{2}(\tau_{C,f}+\tau_{C,r}),
\end{equation}
where $\tau_{C,f}$ and $\tau_{C,r}$ are  the throughput in the thin cloud if using SA$^I$-GS laser and SA$^I$-GS RF link, respectively. $\tau_{C,f}$ and $\tau_{C,r}$ are from Lemma \ref{networkthroughputsaginshaps} and Lemma \ref{thrrfthincloud}.
\end{lemma}

\begin{lemma} \label{networkthroughputsaginshaps}
The average network throughput in the thin cloud using SA$^I$-GS laser link is given as follows:
\begin{equation}
\tau_{C,f} = \frac{D_{C,f}}{T_{C,f}},
\end{equation}
where the average amount of data per transmission slot that SA$^I$ forwards can be given as follows:
\begin{equation} \label{Dablockednamnhunghihi}
    D_{C,f} = [1-P(0)]  P_{SA^I} \cdot X,
\end{equation}
and the average transmission slot length  is given as follows:
\begin{equation} \label{explTforvi}
\begin{split}
T_{C,f} & = P(0) \cdot (1-\rho_{SS}) \frac{X}{C_{SS}} + P(0) \cdot \rho_{SS} \cdot T^o_{SS} \\ & + [1-P(0)] P_{SA^s}  \frac{X}{C_{SS}} + [1-P(0)] P_{SA^s}^{'} \cdot T^o_{SS} \\ & + [1-P(0)]  P_{SA^I} \frac{X}{C^{f,C}_{SG}} + [1-P(0)] P_{SA^I}^{'} \cdot T^o_{SG},
\end{split}
\end{equation}
where $P(0)$ is referred to Lemma \ref{stateprobblockednewvi}.
\end{lemma}
The proof is provided in Appendix \ref{proofnetworkthroughputsaginshaps}.

\begin{lemma} \label{thrrfthincloud}
The average network throughput in the thin cloud using SA$^I$-GS RF link is given as follows:
\begin{equation}
\tau_{C,r} = min([1-P_{C,r}]\beta,\phi),
\end{equation}
where all parameters are referred to Lemma \ref{Acrnamli}, and $P_{C,r}$ is from Lemma \ref{PCrmorning}.
\end{lemma}
\subsection{Rain weather model}
With the rain (SA$^I$-GS laser link), the average queue length $A_{R}$, packet dropping probability $P_R$, and throughput $\tau_R$ can be similarly calculated using Lemmas \ref{theoremaveragequeuehaps}, \ref{pdpblockedvido}, and \ref{networkthroughputsaginshaps}.

\subsection{Foggy weather model}
In foggy weather (SA$^I$-GS RF link), the average queue length $A_{F}$, packet dropping probability $P_F$, and throughput $\tau_F$ can be similarly calculated using Lemmas \ref{Acrnamli}, \ref{PCrmorning}, and \ref{thrrfthincloud}.
\subsection{Combined weather models}
As the satellite moves, it can experience different weather effects. Based on \cite{9655260}, we analyze the system performance under a combined distribution of different weather conditions, including thin cloud, rain, and fog. Let $W_C$, $W_R$, and $W_F$ be the probability of thin cloud, rain, and fog, respectively.
\begin{thm} \label{combinedAt}
   The queue length of SA$^I$ under three weather effects (thin cloud, rain, and fog) combined is given as follows: 
   \begin{equation}
       A_t = W_C \cdot A_C + W_R \cdot A_R + W_F \cdot A_F. 
   \end{equation}
\end{thm}

 We now calculate the computational complexity of the queue length of SA$^I$ under the three weather effects (thin cloud, rain, and fog) considered jointly in Theorem \ref{combinedAt}. The queue length $A_C$ of SA$^I$ under thin cloud conditions is used in Lemma \ref{Actotalmorning}, Lemma \ref{stateprobblockednewvi}, Lemma \ref{theoremaveragequeuehaps}, and Lemma \ref{Acrnamli}. The queue length $A_R$ of SA$^I$ under rain conditions is used in Lemma \ref{stateprobblockednewvi} and Lemma \ref{theoremaveragequeuehaps}. The queue length $A_F$ of SA$^I$ under fog conditions is used in Lemma \ref{Acrnamli}. Based on these lemmas, the computational complexity can be expressed as follows:
\begin{equation} \label{complexneta}
\mathcal{O}(L+O_t).
\end{equation}

\begin{thm} \label{combinedPt}
   The packet dropping probability at SA$^I$ under the thin cloud, rain, and fog combined is given as follows: 
   \begin{equation}
       P_t = W_C \cdot P_C + W_R \cdot P_R + W_F \cdot P_F. 
   \end{equation}
\end{thm}
We calculate the computational complexity of the packet dropping probability at SA$^I$ under the three weather effects (thin cloud, rain, and fog) considered jointly in Theorem \ref{combinedPt} similar to Eq. \eqref{complexneta}. The packet dropping probability at SA$^I$ under thin cloud conditions is used in Lemma \ref{pcquametroi}, Lemma \ref{stateprobblockednewvi}, Lemma \ref{pdpblockedvido}, and Lemma \ref{PCrmorning}; under rain conditions is used in Lemma \ref{stateprobblockednewvi} and Lemma \ref{pdpblockedvido}; and, under fog conditions is used in Lemma \ref{PCrmorning}. 

\begin{thm} \label{combinedtaut}
   The network throughput under three weather effects (thin cloud, rain, and fog) combined is given as follows: 
   \begin{equation}
       \tau_t = W_C \cdot \tau_C + W_R \cdot \tau_R + W_F \cdot \tau_F. 
   \end{equation}
\end{thm}

The network throughput under thin cloud conditions is used in Lemma \ref{taucafternoon}, Lemma \ref{stateprobblockednewvi}, Lemma \ref{networkthroughputsaginshaps}, Lemma \ref{PCrmorning}, and Lemma \ref{thrrfthincloud}; under rain conditions is used in Lemma \ref{stateprobblockednewvi} and Lemma \ref{networkthroughputsaginshaps}; under fog conditions is used in Lemma \ref{PCrmorning} and Lemma \ref{thrrfthincloud}. Based on these lemmas, the computational complexity can be expressed as in Eq. \eqref{complexneta}.

 The proposed method has a linear computational complexity of $\mathcal{O}(L+O_t)$, indicating relatively low computational complexity. Therefore, the method is suitable for future transparent optical architectures, such as the downlink application of the Global Navigation Satellite System Real-Time Network (GNSS-RTN) for autonomous vehicle control, which requires continuous access to accurate geospatial data in real time \cite{10663454}.

\begin{figure*}[ht]
\captionsetup[subfigure]{}
  \centering
  \subfloat[]{\includegraphics[ width=0.29\linewidth]{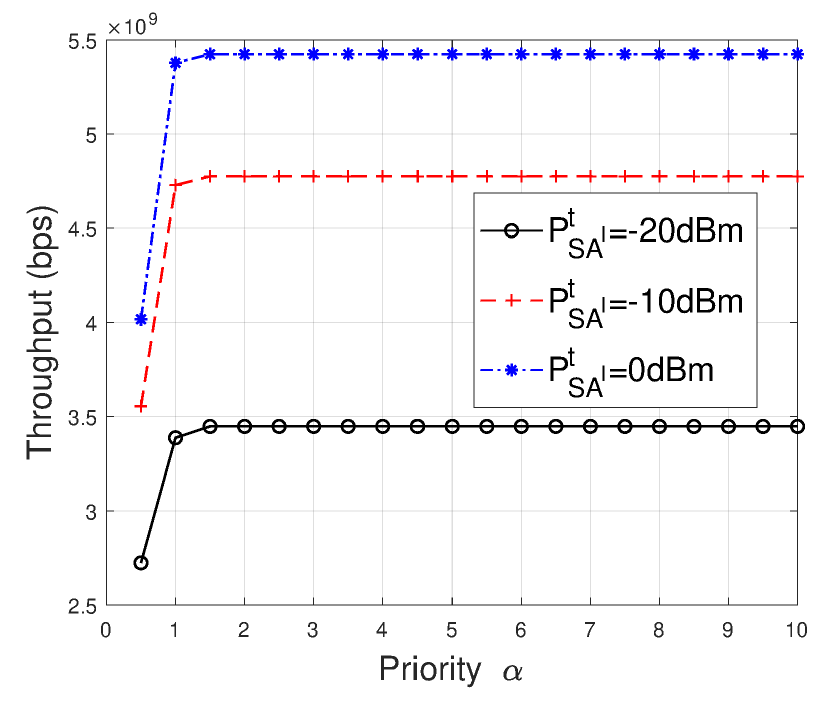}
  \label{throughput_vs_alpha}}
  \subfloat[]{\includegraphics[ width=0.29\linewidth]{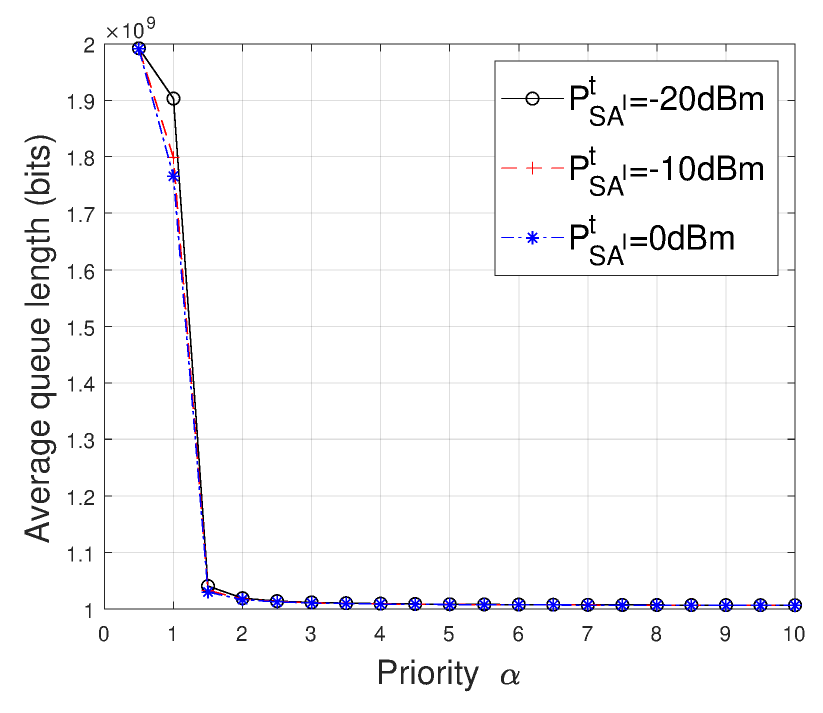}
  \label{queue_vs_alpha}}
  \subfloat[]{\includegraphics[ width=0.29\linewidth]{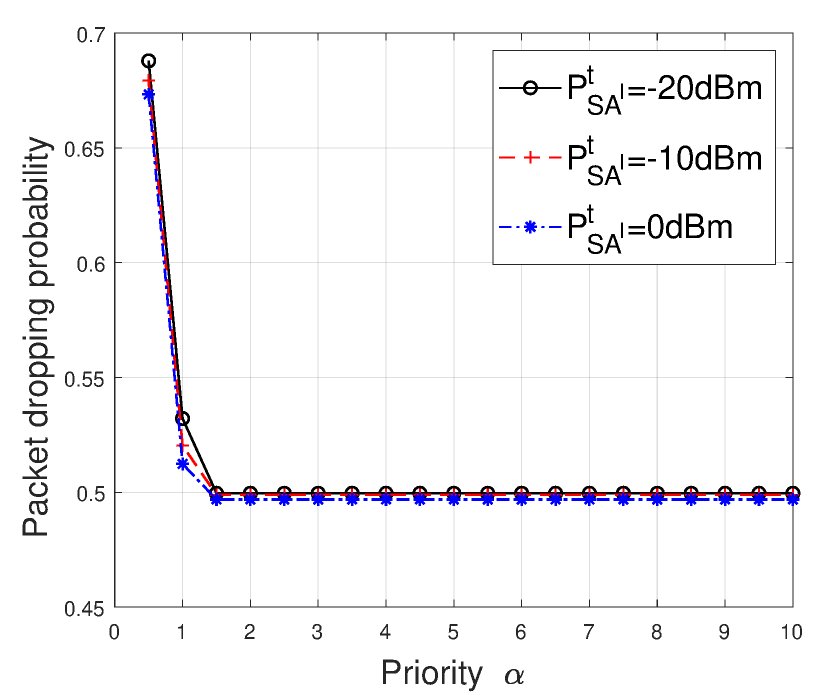}
  \label{dropping_vs_alpha}}
  \caption{Satellite system performance over different SA$^I$ priorities $\alpha$, where the SA$^I$ buffer limit is fixed at $L=100$ packets.}
  \label{analysis_thr_aver_loss_vs_alpha}
  \end{figure*}

    \begin{figure*}[ht]
\captionsetup[subfigure]{}
  \centering
  \subfloat[]{\includegraphics[width=0.29\linewidth]{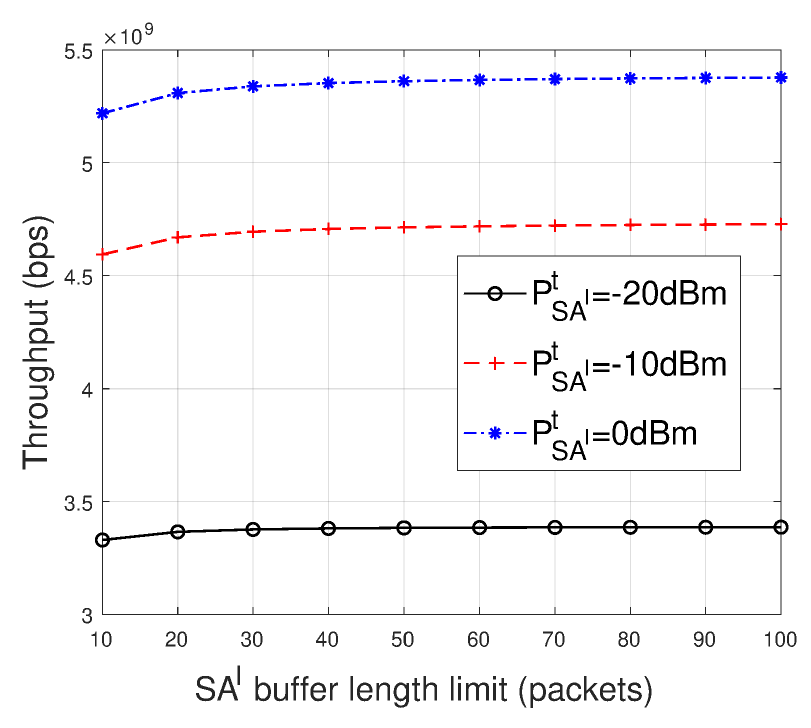}
  \label{throughput_vs_L}}
  \subfloat[]{\includegraphics[ width=0.29\linewidth]{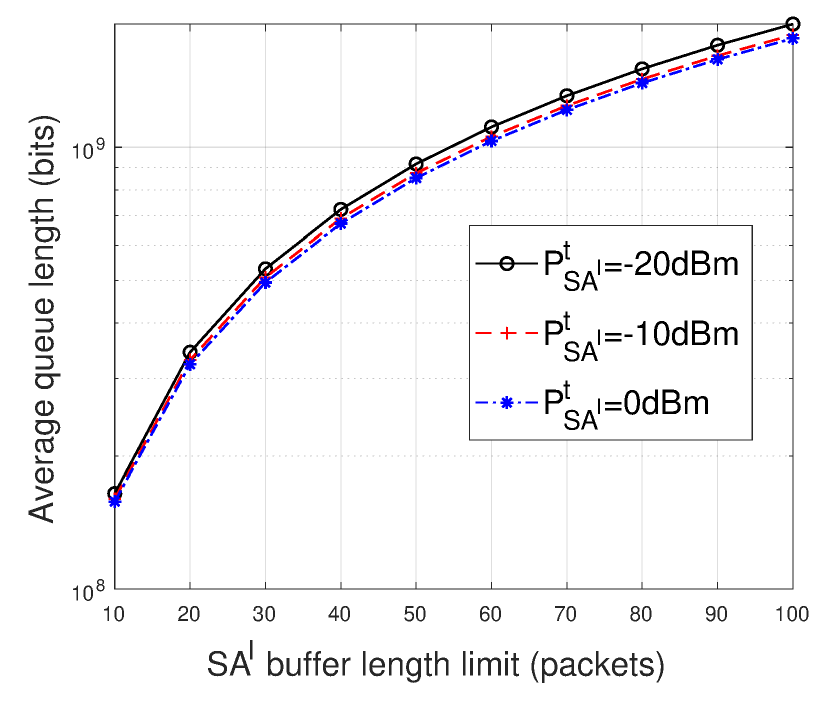}
  \label{queue_vs_L}}
  \subfloat[]{\includegraphics[width=0.29\linewidth]{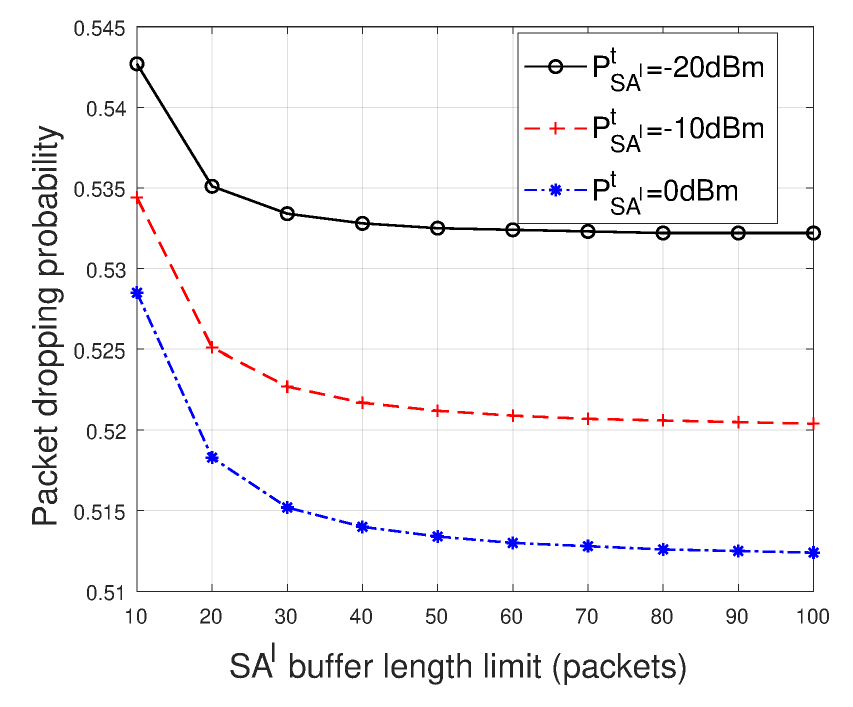}
  \label{dropping_vs_L}}
  \caption{Satellite system performance over different SA$^I$ buffer limits $L$, where the SA$^I$ priority is fixed at $\alpha=1$.}
  \label{analysis_thr_aver_loss_vs_L}
  \end{figure*}
  
\section{Performance evaluation} \label{perforevablockedfnm}
We start the evaluation with the system configuration assumptions. The timeouts $T^o_{SS}$ for SA$_s$-SA$_I$ link and $T^o_{SG}$ for SA$_I$-GS link are $0.01$ s.  The packet size is fixed, i.e., $20$ Mbits. Assume that the average outage probability of SA$^s$-SA$^I$ link is small, e.g., $0.0001$. For the physical properties, we define the parameters for each weather scenario and OGS placement according to \cite{9655260,Samy:23,10673969} and summarize the main parameters used in Table \ref{tab:configuration_parameters}.

\begin{table}[!th]
  \centering
  \caption{Configuration Parameters}
  \label{tab:configuration_parameters}
  \scriptsize
  \setlength{\tabcolsep}{3pt}
  \renewcommand{\arraystretch}{1.02}
  \begin{tabularx}{\columnwidth}{@{}Xr@{}}
    \toprule
    \textbf{Configuration parameter} & \textbf{Value} \\
    \midrule

    \multicolumn{2}{c}{\textbf{LEO satellite and ground station}} \\
    \midrule
    Satellite altitude $H_{\mathrm{sat}}$ & 500 km \\
    Ground-station altitude $H_{\mathrm{GS}}$ & 0.8 km \\
    Zenith angle $\xi$ & $60^{\circ}$ \\
    Elevation angle $\theta_{\mathrm{el}}$ & $30^{\circ}$ \\ 
    Satellite Power $P_{SA^s}^T$ & 0.7~W\\

    \midrule
    \multicolumn{2}{c}{\textbf{FSO link}} \\
    \midrule
    Optical wavelength $\lambda$ & 1550 nm \\
    Optical-to-electrical conversion coefficient $\zeta$ & 0.8 \\
    Receive-aperture radius $a$ & 10 cm \\
    Pointing-jitter standard deviation $\sigma_s$ & 20 cm \\
    Nominal refractive-index parameter $C_0$ (thin cloud) & $5\times10^{-14}$ m$^{-2/3}$ \\
    Nominal refractive-index parameter $C_0$ (rain) & $0.4\times10^{-14}$ m$^{-2/3}$ \\
    Ground wind speed $v_g$ (thin cloud / rain) & 2.8 / 11.176 m/s \\

    \midrule
    \multicolumn{2}{c}{\textbf{RF link}} \\
    \midrule
    RF carrier frequency $f_{\mathrm{RF}}$ & 40 GHz \\
    Transmit antenna gain $G_{\mathrm{T}}$ & 52 dBi \\
    Receive antenna gain $G_{\mathrm{R}}$ & 52 dBi \\
    Oxygen specific attenuation $\omega_{\mathrm{oxy}}$ & 0.1 dB/km \\
    Fog specific attenuation $\omega_{\mathrm{fog}}$ & 339.62 dB/km \\

    \midrule
    \multicolumn{2}{c}{\textbf{Other parameters}} \\
    \midrule
    Noise power $N_0$ & $1.0\times10^{-13}$ W \\
    Outage threshold $\gamma_{\mathrm{th}}$ & 7 dB \\
    State probabilities $(p_0,p_1,p_2)$ & $(1/3,1/3,1/3)$ \\
    \bottomrule
  \end{tabularx}
\end{table}

 For fairness, assume that the probabilities of the different weather types ($W_C$, $W_R$, and $W_F$) are equal \cite{9655260}. The observed time $O_t$ in case of fog, it only needs a few dozen of seconds to get convergent results which were also proved in \cite{11124522}, while the other weather effects may be have a longer duration.


In Fig. \ref{analysis_thr_aver_loss_vs_alpha}, we consider a SA$^I$ buffer with $L=100$ packets, different SA$^I$ transmission scheduling priorities $\alpha = [0.5,10]$ in step of $0.5$ ($\alpha$ is only used for SA$^I$-GS laser), and transmission power  of SA$^I$ $P^t_{SA^I}=\{-20,-10,0\}$ dBm.  We consider $\alpha=1$ as the baseline scenario, in which neither the intermediate satellite nor the source satellite is given priority, resulting in an equal transmission scheduling priority between them. As we analyze results under combined weather effects. The smaller the $P^t_{SA^I}$ value is, the worse the performance is because the probability of SA$^I$-GS link outage increases with decreasing $P^t_{SA^I}$. Also, the throughput in Fig. \ref{throughput_vs_alpha} increases with increasing $\alpha$. This is because the buffered packets can be removed faster with a higher $\alpha$ (average queue length in Fig. \ref{queue_vs_alpha} and packet dropping probability in Fig. \ref{dropping_vs_alpha} decrease with increasing $\alpha$). At a fixed transmission power $P^t_{SA^I}$, we can increase the performance by adjusting the priority $\alpha$ without needing an increment of $P^t_{SA^I}$. However, the throughput achieves a maximum at $\alpha \geq 1.5$ and configuring $\alpha$ larger than $1.5$ is unnecessary because the packet dropping probability at $\alpha \geq 1.5$ is constant. The reason is that setting a SA$^I$ priority $\alpha$ higher than $1.5$ will occupy more redundant transmission chance over SA$^s$, leading to sometime less data transmitted from SA$^s$ due to using an inefficient frequency resource. The analytical results show that the system achieves better performance at $\alpha=1.5$ than at the baseline value of $\alpha=1$.
  
In Fig. \ref{analysis_thr_aver_loss_vs_alpha}, when SA$^I$-GS RF link is used, the performance does not depend on the SA$^I$ priority $\alpha$. The performance is constant after observed time $O_t$ because the arrival data rate of SA$^s$-SA$^I$ laser link is much larger than the released data rate of SA$^I$-GS RF link. Thus, the queue length of SA$^I$ buffer is always approximately full, and the packet dropping probability is only the ratio of the amount of released data to the amount of arrival data. Hence, the variant performance over different $\alpha$ values mainly comes from the case of the weather using SA$^I$-GS laser link. Due to after $\alpha \geq 1.5$, the performance is constant. Thus, to get a much better quality of service, we need to sacrifice the consumed energy of satellite by increasing $P^t_{SA^I}$. In addition, the average queue length and packet dropping probability are relatively high, even increasing $P^t_{SA^I}$ and $\alpha$. This is due to bad performance by using SA$^I$-GS RF link which is independent with using the priority $\alpha$. Thus, the satellite systems prioritize using the laser technology, while the RF technology is used as the backup \cite{11124522}. 

We now perform a sensitivity analysis of the weather distribution and its impact on system performance. The results show that system performance improves when weather conditions that favor laser links dominate. For example, rainy conditions are favorable for the use of laser links. We consider a rain-weather scenario in which the SA$^I$-GS laser link is used and vary the priority parameter $\alpha$. The system achieves its maximum performance at $\alpha = 1.5$. Specifically, the throughput reaches $3.8768$ Gbps, $4.9615$ Gbps, and $6.1077$ Gbps for $P^t_{SA^I}=-20$ dBm, $P^t_{SA^I}=-10$ dBm, and $P^t_{SA^I}=0$ dBm, respectively. The queue length is minimized to $0.0936$ Gbits, $0.0740$ Gbits, and $0.0647$ Gbits, respectively. Moreover, the packet-dropping probability is zero for all three transmit-power settings. With the analytical results, we see that the system performance also depends strongly on the distribution of the different weather conditions. For example, when the probabilities of the different weather conditions are equal, the packet-dropping probability is close to 50$\%$ as shown in Fig. \ref{analysis_thr_aver_loss_vs_alpha}, whereas when rainy conditions completely dominate the weather distribution, the packet-dropping probability approaches $0$. These numerical results demonstrate that the weather distribution has a significant impact on the performance of the satellite communication system.


  Next, we consider a SA$^I$ buffer with $L=[10,100]$ packets in step of $10$, SA$^I$ priority $\alpha=1$, and $P^t_{SA^I}=\{-20,-10,0\}$ dBm. The throughput shown (Fig. \ref{throughput_vs_L}) increases with increasing $L$ because the SA$^I$ buffer with a higher $L$ value can save more data which is proved via the average queue length (Fig. \ref{queue_vs_L}), so this leads to lower packet dropping probabilities (Fig. \ref{dropping_vs_L}). However, the total satellite system performance slowly increases with increasing $L$ because of the inefficiency in case of the weather using SA$^I$-GS RF link as analyzed above. 

  
\section{Conclusion} \label{conclpaper}
We proposed an analytical framework based on Markov chain model to calculate the throughput and on-board buffering requirements for hybrid RF/FSO networks. The analytical results also show that instead of increasing transmission power, which increases energy and weight requirements,  a suitable optical transmission scheduling priority can be used to maximize throughput and minimize electronic buffer size. We observed best results when using laser downlinks, resulting in a small buffer and zero packet dropping probability.  The proposed analytical model has several limitations that can be addressed in future work. First, the framework can be extended to multiple sources and ground stations to better represent realistic LEO networks. Second, switching overhead between RF and optical links needs to be considered, which is critical. Third, we need a model of variable packet size.

\section*{Acknowledgment}
 This work is partially supported by the Federal Ministry of Research, Technology, and Space (BMFTR)  xG-RIC project as part of the research program Communication Systems  "Souver\"an. Digital. Vernetzt."  (grant number 16KIS2429K).
\appendix
\subsubsection{Proof of Lemma \ref{fiablockednew}} \label{prooffiablockednew}
As the state space in the Markov chain has $L$ states, it is finite. In addition, this Markov chain is irreducible because we observe Fig. \ref{Markov_block_prob}  that any state $i \in [0,L]$ can be reachable from any other state $j \in [0,L] \setminus i$ with a non-zero probability. Finally, we prove that if any state from this Markov chain is aperiodic, then it is aperiodic. Any state $i$ with the period $d (i )$ has the greatest common denominator from all integer values larger than $0$, whereby there exists a transition probability larger than $0$ from any state $i$ traversing to the other states, and then going back to itself. In this Markov chain, there always exists a state $i$ with a non-zero self-transition probability, $d(i)=1$. Thus, the state $i$ is aperiodic.

\subsubsection{Proof of Lemma \ref{stateprobblockednewvi}} \label{proofstateprobblockednewvi}
In Fig. \ref{Markov_block_prob}, the balance equations can be given as follows:
\begin{equation} \label{baleqblocked1}
P(1) = \frac{1-\rho_{SS}}{P_{SA^I} } P(0); P(i+1) = \frac{ P_{SA^s} }{ P_{SA^I} } P(i), \forall i \in [1,L).
\end{equation}
With Eqs. \eqref{baleqblocked1}, and $\sum_{i=0}^L P(i)=1$, we get Lemma \ref{stateprobblockednewvi}.

\subsubsection{Proof of Lemma \ref{networkthroughputsaginshaps}} \label{proofnetworkthroughputsaginshaps}

Based on Fig. \ref{Markov_block_prob}, if the buffer is empty with the state probability $P(0)$, the transmission probability of SA$^s$ without competing with SA$^I$ that sends successfully one packet to SA$^I$ is $1-\rho_{SS}$ with the transmission slot length $\frac{X}{C_{SS}}$, i.e., $P(0) \cdot (1-\rho_{SS}) \frac{X}{C_{SS}}$. In case of empty buffer, the timeout time $T^o_{SS}$ for the SA$^s$-SA$^I$ laser link due to its outage probability $\rho_{SS}$ causing packets unsuccessfully sent is $P(0) \cdot \rho_{SS} \cdot T^o_{SS}$. If the buffer is nonempty with the probability $P(\bar{0})=1-P(0)$, the transmission probability of SA$^s$ with competing with SA$^I$ that sends successfully one to SA$^I$ is $P_{SA^s}$ with the transmission slot length $\frac{X}{C_{SS}}$, i.e., $P( \bar{0}) \cdot P_{SA^s}  \frac{X}{C_{SS}}$. Also, in case of nonempty buffer, the transmission probability of SA$^s$ with competing with SA$^I$ that sends unsuccessfully one to SA$^I$ is $P_{SA^s}^{'}$ with timeout time $T^o_{SS}$, i.e., $P( \bar{0}) \cdot P_{SA^s}^{'} \cdot T^o_{SS}$. Similarly as explained above, we have the transmission slot length if SA$^I$ wins transmission chances, i.e., $P( \bar{0}) \cdot P_{SA^I} \frac{X}{C^{f,C}_{SG}} + P( \bar{0}) \cdot P_{SA^I}^{'} \cdot T^o_{SG}$. So, we get Eq. \eqref{explTforvi}. If the buffer is nonempty, the transmission probability of SA$^I$ with competing with SA$^s$ that sends successfully one packet ($X$ bits) to GS is $P_{SA^I}$, i.e., Eq. \eqref{Dablockednamnhunghihi}.

\bibliographystyle{IEEEtran}

\bibliography{nc-rest}

\end{document}